\documentclass[eqsecnum,showpacs,aps]{revtex4}
\usepackage{graphicx}

\begin{document}

\draft


\title{Nonradial oscillations of quark stars}

\author{
Hajime Sotani}
\email{sotani@gravity.phys.waseda.ac.jp}

\author{
Tomohiro Harada}
\email{harada@gravity.phys.waseda.ac.jp}

\address{
Department of Physics, Waseda University,
Okubo 3-4-1, Shinjuku, Tokyo 169-8555, Japan
}

\date{\today}

\begin{abstract}
Recently, it has been reported that a candidate for a 
quark star may have been observed.
In this article, we pay attention to quark stars with 
radiation radii in the reported range.
We calculate nonradial oscillations of 
$f$-, $w$- and $w_{\rm II}$-modes.
Then,
we find that the dependence of the $f$-mode
quasi-normal frequency on the bag constant and stellar radiation radius 
is very strong and different from that
of  the lowest $w_{\rm II}$-mode quasi-normal frequency. 
Furthermore we deduce a new empirical formula
between the $f$-mode frequency of gravitational waves 
and the parameter of 
the equation of state for quark stars.
The observation of gravitational waves both of the $f$-mode
and of the lowest $w_{\rm II}$-mode would provide a powerful 
probe for the equation of state of quark matter and 
the properties of quark stars. 
\end{abstract}




\maketitle


\section{Introduction}

Since the 1980s, there have been many studies about objects known as 
quark stars. If the true ground state of matter is not the nuclear state 
but a quark state as Witten conjectured \cite{Witten1984}, 
there may exist objects composed of quark matter.          
For simplicity, the bag model equation of state (EOS) has often been 
used for studying quark matter. 
Lattimer and Prakash calculated stellar properties with 
this EOS \cite{Lattimer2001}. In reality, it is quite difficult to 
determine the EOS of quark matter by experiments on the ground, 
because quark matter will appear at extremely high density. 
In this respect, the observation of astrophysical compact objects
may be a unique means to determine the equation of state for 
cold high-density matter \cite{Lindblom1992,Harada2001}. 

Recently, the radiation radius of the compact star 
RXJ185635-3754 was estimated to be very small ($3.8 \mbox{km}
\alt R_{\infty} \alt 8.2 \mbox{km}$) 
based on X-ray observations 
by {\em Chandra} \cite{Drake2002}. 
The radiation radius is related as $R_{\infty}\equiv R/\sqrt{1-2M/R}$ 
to the areal radius $R$ and mass $M$ of the star. 
This compact star is too small to construct with  
the normal EOS's which are currently adopted for neutron stars \cite{Lattimer2001}. 
This star should be a candidate for a quark star. 
Inspired by this observation, 
Nakamura proposed several scenarios about the formation 
of quark stars \cite{Nakamura2002}. 

However, it should be noted that 
there exist many subsequent articles which 
question the whole observation.
For example, Pons et al. pointed out, by employing the data of 
not only X-ray observation but also UV/optical observation, 
that if RXJ185635-3754 is represented by 
the simplest uniform-temperature heavy-element 
atmospheric model, this compact star has $R_{\infty}\approx 8$km, 
$M\approx 0.9M_{\odot}$ and $R\approx 6$km \cite{Pons2002}. 
Though this radiation radius is included in 
the range suggested by \cite{Drake2002}, 
these properties can not be constructed, 
even if the quark matter EOS is used. 

If a compact object oscillates for some reason, 
gravitational waves are emitted from the object. 
The trigger of the oscillation may be a nonspherical supernova 
explosion, coalescence with another compact 
object, violent mass influx 
due to possible instability of the surrounding accretion disk, and so on.
The oscillations damp out because the gravitational waves carry away 
oscillational energy. So these oscillations are called quasi-normal 
modes (QNMs). If the gravitational waves from a compact object are 
detected, we can get information about the source object. 

Gravitational wave observation projects, such as LIGO \cite{Althouse1992}, 
VIRGO \cite{Giazotto1990}, GEO600 \cite{Hough1996} and 
TAMA300 \cite{Tsubono1995}, are making remarkable progress in sensitivity. 
Among them, TAMA300, GEO600 and LIGO are now in operation.
If highly sensitive gravitational wave detectors are 
available, we may obtain a large amount of data for frequencies and 
damping times of gravitational waves emitted from compact objects.
The systematic study of gravitational wave modes 
leads to the so called ``gravitational wave asteroseismology'',
which has been initiated and being presented by various authors
(see \cite{Andersson1996,Andersson1998,Kokkotas1999,
KokkotasAndersson2001,Kokkotas2001}).
If we have a good empirical formula for gravitational wave oscillational
frequencies and damping times as functions of stellar properties, 
it will be very useful
in obtaining information for source stars from gravitational
wave observations.
In this context,
Andersson and Kokkotas calculated QNMs with various EOSs and 
proposed an empirical formula between the QNMs and properties 
of neutron stars \cite{Andersson1996,Andersson1998}. 
Kokkotas, Apostolatos and Andersson 
improved this empirical formula by one which included the relevant statistical errors 
\cite{Kokkotas2001}. 
These works treated the polar modes of oscillation and 
the results are well summarized in \cite{Kokkotas1999,KokkotasAndersson2001}. 
Benhar, Berti and Ferrari calculated axial modes and showed that 
the axial mode gives more direct and explicit information 
on the stiffness of EOS compared to the polar mode \cite{Benhar2000}. 
All these works are for normal neutron stars, i.e., assuming 
normal equations of state for nuclear matter.

On the other hand, Yip, Chu and Leung studied 
nonradial stellar oscillations of quark stars whose radii are around 
10 km \cite{Yip1999}. However, the reported radius of the star 
RXJ185635-3754 is much smaller than the radii of the stellar models 
examined by Yip, Chu and Leung.
In this paper, we calculate nonradial oscillations of quark stars 
whose radiation radii 
are in the range of $3.8 \mbox{km} \alt R_{\infty} \alt 8.2 \mbox{km}$, 
and examine the relation 
between the QNMs of quark stars and the EOS of quark matter.
Our main concern in this article is whether one could distinguish the 
EOS of quark matter by the direct detection of the gravitational wave.

QNMs are classified into two classes. One is a class of fluid modes, which 
mainly couple with the stellar fluid, while the other is one of 
spacetime modes which are 
connected with spacetime oscillation, 
which couple mainly to metric perturbations.
Non-rotating stars admit fluid modes only for polar perturbations. 
The most widely studied fluid modes
are $f$-, $p$- and $g$-modes \cite{Andersson1998,Sotani2001}. The 
$f$-mode is the fundamental mode. There is only one $f$-mode for each 
$l$. The $p$-mode is the pressure mode which arises from fluid pressure. 
The $g$-mode is the gravity mode which is caused by buoyancy 
due to density discontinuity and/or the temperature gradient of stars and other factors. 
The damping rates of these fluid modes are very small compared 
with the frequency.
The spacetime modes include
$w$- and $w_{\rm II}$- modes \cite{Kokkotas1992,Leins1993}.
The damping rate of $w$-modes is comparable to the frequency.
For each stellar model, there are found one or a few $w_{\rm II}$-modes, 
whose complex frequency is located near the imaginary axis.
Here we deal with only polar perturbation because 
our main concern is the relationship between the gravitational waves and EOS.
For simplicity, we pay attention here only to $f$-, $w$- and $w_{\rm II}$-modes.
Recently, Kojima and Sakata 
calculated the $f$-mode quasi-normal frequency for quark star
models, and pointed out the possibility of distinguishing between quark stars 
and neutron stars by detecting both the $f$-mode
frequency and damping rate \cite{Kojima2002}.

The plan of this paper is as follows. We present basic equations to 
construct quark stars and show the properties 
of quark star structure
in Sec. \ref{sec:QuarkStar}. 
In Sec. \ref{sec:Oscillation},
we present the equations and method to determine the 
QNMs for spherically symmetric stars,
show numerical results for the quasi-normal frequencies of quark-star models 
constructed in Sec. \ref{sec:QuarkStar},
and discuss their implications. 
Moreover, we apply the empirical formula proposed by Kokkkotas et al. 
to quark star models and 
deduce a new empirical formula from our numerical results 
between the $f$-mode frequency of the gravitational wave and 
the parameter of EOS
in Sec. \ref{sec:empiricalrelation}.
We conclude in Sec. \ref{sec:conclusion}. 
We adopt the unit of $c=G=1$, 
where $c$ and $G$ denote the
speed of light and gravitational constant, respectively, and the metric
signature of $(-,+,+,+)$.

\section{Structure of Quark Stars}
\label{sec:QuarkStar}

We consider a static and spherical star. For this case, the metric 
is described by
  \begin{equation}
  ds^{2}=-e^{2\Phi}dt^{2}+e^{2\Lambda}dr^{2}+r^{2}\left(d\theta^{2}+\sin^{2}
           \theta d\phi^{2}\right),
  \label{metric_star}
  \end{equation}
where $\Phi,\Lambda$ are metric functions of $r$. The mass function 
$m(r)$ is defined as
   \begin{equation}
    m(r) = \frac{1}{2}r\left(1-e^{-2\Lambda}\right). \label{ephi}
   \end{equation}
This mass function satisfies
  \begin{equation}
   \frac{dm}{dr}    = 4\pi r^2\rho, \label{dmdr}
  \end{equation}
where $\rho$ is the energy density.
The equilibrium stellar model is derived by solving the 
Tolman-Oppenheimer-Volkoff equation,
  \begin{equation}
    \frac{dP}{dr}   = -\frac{\left(\rho+P\right)\left(m+4\pi r^3P\right)}
                         {r\left(r-2m\right)}, \label{dpdr}
  \end{equation}
where $P$ is the pressure of the fluid, and the potential $\Phi$ is 
given by
  \begin{equation}
    \frac{d\Phi}{dr}   = \frac{\left(m+4\pi r^3P\right)}
                         {r\left(r-2m\right)}. \label{dphidr}
  \end{equation}
For solving these equations, we need an additional equation, i.e., the 
equation of state.

Though a quark star may be either a light quark star made up of $u$ and 
$d$ quarks or a strange quark star composed of a mixture of $u$, 
$d$ and $s$ quarks, we construct simple models by using the 
bag model EOS, which neglects the masses of $u$ and $d$ quarks. 
This bag model EOS is characterized by three parameters: the 
strong interaction coupling constant $\alpha_c$, 
the bag constant $B$ and the mass $m_{s}$ of the $s$ quark. 
The dependence of stellar properties on the bag constant is much 
stronger than that on $\alpha_c$ and $m_{s}$ \cite{Kohri2002}.
In this paper we use the bag model EOS derived for a massless strange quark,
\begin{equation}
  P=\frac{1}{3}(\rho-4B). \label{bagmodel}
\end{equation}
The bag constant $B$ is a positive energy density, which 
corresponds to latent heat. The stellar radius $R$ is defined as
the position 
where the pressure is zero. At the surface, the 
density profile is discontinuous. 
Although the value of the bag constant, which is a 
phenomenological parameter, should be determined by the underlying strong 
interaction dynamics, it is difficult to determine this value from
our present understanding of QCD. However, if strange quark  matter 
is the true ground state,
the energy of this true ground state per particle would not be over 
the nucleon mass 939 MeV at baryon density for zero pressure 
matter. This constraint implies a maximum value of the bag constant 
$B$ as follows \cite{Prakash1990}:
\begin{equation}
  B\le B_{\text{max}}=
94.92\left(1-\frac{2\alpha_c}{\pi}\right) \text{MeV fm}^{-3}.
\end{equation}

\begin{center}
\begin{figure}[htbp]
\includegraphics{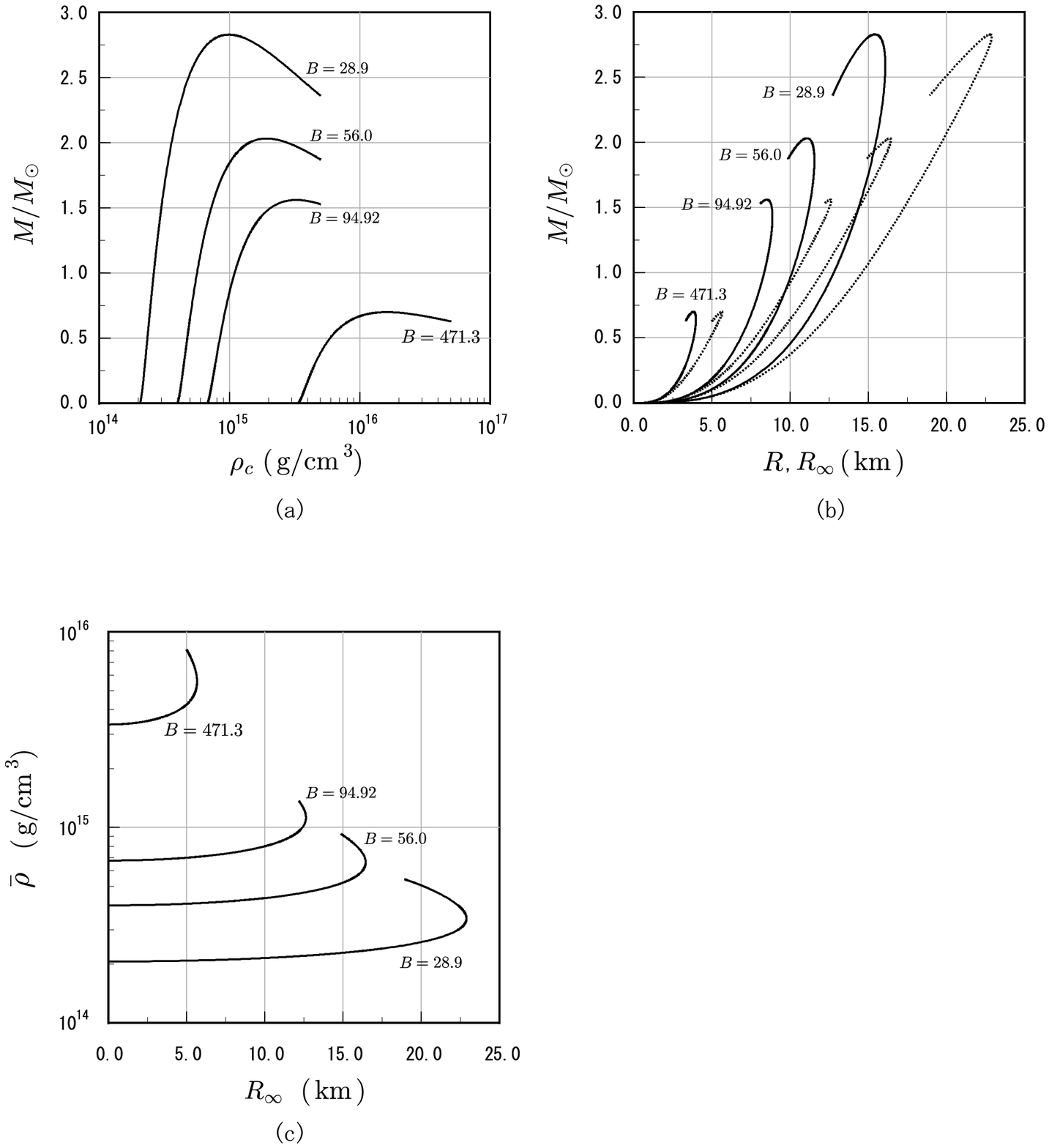}
\caption{Relations (a) between mass $M$ and central density $\rho_c$, 
(b) between $M$ and 
radius $R$ (solid line)
and between $M$ and radiation radius $R_{\infty}$
(broken line),
and (c) between ``average 
density'' $\bar{\rho}\equiv 3M/4\pi R^3$ and radiation radius $R_{\infty}$
of quark stars 
for four values of the bag constant, $B=28.9\mbox{MeV fm}^{-3}$, 
$56.0\mbox{MeV fm}^{-3}$, 
$94.92\mbox{MeV fm}^{-3}$ and $471.3 \mbox{MeV fm}^{-3}$.}
\label{fig_property}
\end{figure}
\end{center}

For definiteness, we use three values of the bag constant, $B=28.9$, 
$56.0$ and $94.92$ MeV fm$^{-3}$. Although
$B_{st}^{1/4}=145$ MeV or $B_{st}=57.8$ MeV fm$^{-3}$ has been 
often used for the 
study of quark-gluon plasma so far, we adopt $B=56.0$ MeV 
fm$^{-3}$ to compare with the results in \cite{Yip1999}. The other 
values of the bag constant are the maximum value for $\alpha_{c}=0$
and half the standard
value $B_{st}$. We show the mass $M$ of quark stars as a function of 
the central density $\rho_c$ for these three values of the bag 
constant in Fig. \ref{fig_property}(a), the relations between $M$ 
and $R$ and between 
$M$ and $R_{\infty}$ in 
Fig. \ref{fig_property}(b), and the relation between ``average density''
$\bar{\rho}\equiv 3M/4\pi R^{3}$ and $R_{\infty}$ in 
Fig. \ref{fig_property}(c). These figures are plotted for $4B < \rho_c 
\le 5.0\times 10^{15}$ g/cm$^{-3}$. The radiation radius may be 
determined from X-ray observations such as \cite{Drake2002}.
We pay attention to the stars whose radiation radii are 
in the range of $3.8 - 8.2$ km. In this range of radiation 
radius, we pick up three values, $R_{\infty} = 3.8$, $6.0$ and $8.2$ km.

On the other hand, Nakamura argued that the mass of the compact star 
reported in \cite{Drake2002} should be roughly $0.7 M_{\odot}$ 
in order to account for its observed  X-ray
luminosity \cite{Nakamura2002}. 
If we use a value smaller than
$B_{\text{max}}$, however, the mass of a quark star whose radiation radius is 
in the range of $3.8 - 8.2$ km is far below $0.7 M_{\odot}$ as
seen in Fig. \ref{fig_property}(b). Therefore he pointed out 
the possibility of adopting the value $B=471.3$ MeV 
fm$^{-3}$, which is rather greater than $B_{\text{max}}$. 
We consider a quark-star mass of $M=0.7M_{\odot}$ 
for this value of the bag constant
to compare with other models. Also for the case 
$B=471.3$ MeV fm$^{-3}$, we plot the relation between
$M$ and $\rho_c$ in Fig. \ref{fig_property}(a), the 
relations between $M$ and $R$ and 
between $M$ and $R_{\infty}$ in Fig. \ref{fig_property}(b), 
and the relation 
between $\bar{\rho}$ and 
$R_{\infty}$ in Fig. \ref{fig_property}(c). In this case, each plot
is for $4B < \rho_c \le 5.0\times 10^{16}$ g/cm$^{-3}$.
The properties of our quark-star models 
are tabulated in Table \ref{tab_property_QS}.

\section{Nonradial Oscillations of Quark Stars}
\label{sec:Oscillation}
\subsection{Method}
The QNMs are 
determined by solving the perturbation equations with appropriate 
boundary conditions. The metric perturbation is given by
   \begin{equation}
    g_{\mu\nu} = g^{(B)}_{\mu\nu}+h_{\mu\nu},
   \end{equation}
where $g^{(B)}_{\mu\nu}$ is the background metric of a spherically 
symmetric star (\ref{metric_star}). We have applied a formalism 
developed by Lindblom and Detweiler \cite{Lindblom1985} for 
relativistic nonradial stellar oscillations. In this formalism, 
$h_{\mu\nu}$ for polar perturbations is described as
   \begin{equation}
    h_{\mu\nu} = \left(
    \begin{array}{cccc}
    r^l\hat{H}e^{2\Phi} & i\omega r^{l+1}\hat{H}_{1} & 0 & 0  \\
    i\omega r^{l+1}\hat{H}_{1} & r^l\hat{H}e^{2\Lambda} & 0 & 0  \\
    0 & 0 & r^{l+2}\hat{K} & 0 \\
    0 & 0 & 0              & r^{l+2}\hat{K}\sin^{2}\theta
   \end{array}
   \right) Y^{l}_{m}\,e^{i\omega t},
   \end{equation}
where $\hat{H},\hat{H}_{1},$ and $\hat{K}$ are perturbed metric
functions with respect to $r$, and the components of the Lagrangian 
displacement of fluid perturbations are expanded as
   \begin{eqnarray}
    \xi^{r}      &=& \frac{r^l}{r}e^{\Lambda}\hat{W}Y^{l}_{m}\,
                     e^{i\omega t}, \\
    \xi^{\theta} &=& -\frac{r^l}{r^2}e^{\Lambda}\hat{V}
                     \frac{\partial}{\partial \theta}
                     Y^{l}_{m}\,e^{i\omega t}, \\
    \xi^{\phi}   &=& -\frac{r^l}{r^2\sin^2\theta}e^{\Lambda}\hat{V}
                     \frac{\partial}{\partial \phi}Y^{l}_{m}\,e^{i\omega t},
   \end{eqnarray}
where $\hat{W} $ and $ \hat{V}$ are  functions of $r$.

Then the perturbation equations derived from Einstein equations are given by
   \begin{eqnarray}
    &&\frac{d\hat{H}_{1}}{dr}=-\frac{1}{r}
     \left[l+1+\frac{2m}{r}e^{2\Lambda}+4\pi r^2(P-\rho)e^{2\Lambda}\right]
     \hat{H}_{1} \nonumber \\
    &&\hspace{1.2cm}+\frac{1}{r}e^{2\Lambda}
     \left[\hat{H}+\hat{K}+16\pi (P+\rho)\hat{V}\right],\label{perturbation1} \\
    &&\frac{d\hat{K}}{dr}=\frac{l(l+1)}{2r}\hat{H}_{1}
     +\frac{1}{r}\hat{H}
     -\left(\frac{l+1}{r}-\frac{d\Phi}{dr}\right)\hat{K}
     +\frac{8\pi}{r}(P+\rho)e^{\Lambda}\hat{W}, \label{perturbation2} \\
    &&\frac{d\hat{W}}{dr}=-\frac{l+1}{r}\hat{W}+re^{\Lambda}
     \left[\frac{1}{\gamma P}e^{-\Phi}\hat{X}-\frac{l(l+1)}{r^{2}}\hat{V}
     -\frac{1}{2}\hat{H}-\hat{K}\right], \label{perturbation3} \\
    &&\frac{d\hat{X}}{dr}=-\frac{l}{r}\hat{X}+(P+\rho)e^{\Phi}
     \biggl[\frac{1}{2}\left(\frac{d\Phi}{dr}-\frac{1}{r}\right)\hat{H}
     -\frac{1}{2}\left(\omega^{2}re^{-2\Phi}+\frac{l(l+1)}{2r}\right)
     \hat{H}_{1}  \nonumber \\
    &&\hspace{1.2cm}+\left(\frac{1}{2r}-\frac{3}{2}\frac{d\Phi}{dr}\right)
     \hat{K}-\frac{l(l+1)}{r^{2}}\frac{d\Phi}{dr}\hat{V}
     \nonumber \\
    &&\hspace{1.2cm}-\frac{1}{r}
     \left(\omega^{2}e^{-2\Phi+\Lambda}+4\pi(P+\rho)e^{\Lambda}
     -r^{2}\left\{\frac{d}{dr}\left(\frac{1}{r^{2}}e^{-\Lambda}\frac{d\Phi}{dr}
     \right)\right\}\right)\hat{W}\biggr], \label{perturbation4} \\
    &&\biggl[1-\frac{3m}{r}-\frac{l(l+1)}{2}-4\pi r^{2}P\biggr]\hat{H}
     -8\pi r^{2}e^{-\Phi}\hat{X} \nonumber \\
    &&\hspace{1.2cm}+r^{2}e^{-2\Lambda}\left[\omega^{2}e^{-2\Phi}
     -\frac{l(l+1)}{2r}\frac{d\Phi}{dr}\right]\hat{H}_{1}
     \nonumber \\
    &&\hspace{1.2cm}-\left[1+\omega^{2}r^{2}e^{-2\Phi}-\frac{l(l+1)}{2}
     -\left(r-3m-4\pi r^{3}P\right)\frac{d\Phi}{dr}\right]\hat{K}=0,
     \label{perturbation5} \\
    &&\hat{V}=\frac{e^{2\Phi}}{\omega^{2}(P+\rho)}
     \left[e^{-\Phi}\hat{X}
     +\frac{1}{r}\frac{dP}{dr}e^{-\Lambda}\hat{W}+\frac{1}{2}(P+\rho)\hat{H}
     \right] .
\label{perturbation6}
   \end{eqnarray}
In deriving the above equations for perturbations, we assume a perfect fluid.
Rigorously speaking, it will not be valid because of the existence of matter viscosity.
However, little is known about the viscosity of cold quark matter. Here we omit the viscosity
as a possible approximation.
In Eq. (\ref{perturbation3}), $\gamma$ is the adiabatic index of the unperturbed stellar 
model, which is calculated as
\begin{equation}
\gamma=\frac{\rho+P}{P}\frac{dP}{d\rho}=\frac{1}{3}\frac{\rho+P}{P},
\end{equation}
where the  bag model EOS (\ref{bagmodel}) is used in the second
equality.
The set of Eqs. (\ref{perturbation1}) $-$ (\ref{perturbation4}) is a 
set of differential equations connecting the variables $\hat{H}_{1}$, 
$\hat{K}$, $\hat{W}$ and $\hat{X}$, and Eqs. (\ref{perturbation5}) and 
(\ref{perturbation6}) are the algebraic equations for the variables 
$\hat{H}$ and $\hat{V}$. The perturbation equations outside the star 
are described by the Zerilli equations. By imposing boundary 
conditions such that perturbative variables are regular at the center of 
the star, the Lagrangian perturbation of pressure vanishes at the 
stellar surface, and the gravitational wave is only an outgoing one at 
infinity, one can reduce this to an eigenvalue problem. The 
boundary condition at the stellar surface is $\hat{X}=0$, 
because $\hat{X} \equiv -r^{-l} e^{\Phi} \Delta P$, where $\Delta P$ 
is Lagrangian perturbation of pressure. Furthermore, we set the term 
$e^{-\Phi} \hat{X}/\gamma P$ in Eq. (\ref{perturbation3}) to zero at 
the stellar surface, because $\gamma P = 4B/3$ at $r = 
R$. For the treatment of the boundary condition at infinity, we adopt 
the method of continued-fraction expansion proposed by 
Leaver \cite{Leaver1985}. The details of the determination of quasi-normal
frequencies are given in \cite{Sotani2001}.

\subsection{Numerical Results}
\label{sec:results}

\subsubsection{$f$-mode}

\begin{center}
\begin{figure}[htbp]
\includegraphics{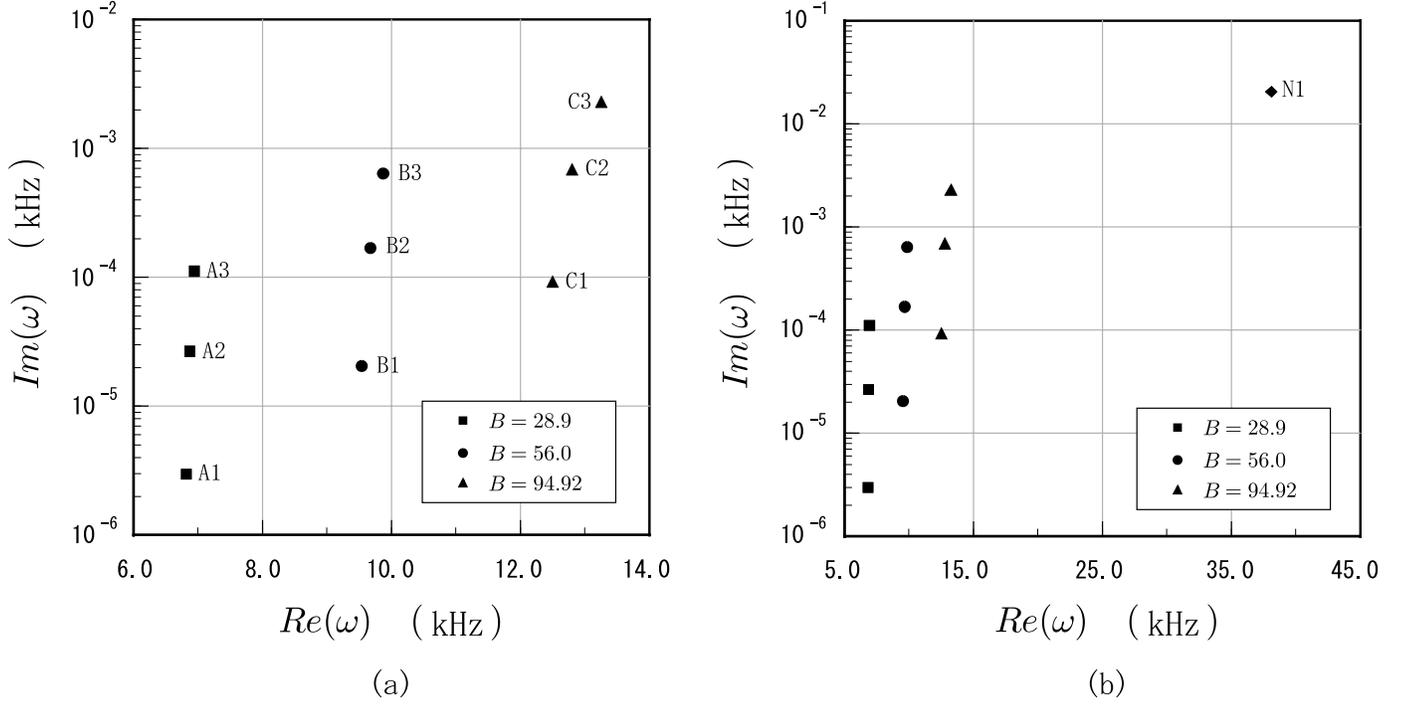}
\caption{Complex frequencies of $f$-mode for quark stars. 
The squares, triangles and circles correspond to 
stars with $B=28.9$ MeV fm$^{-3}$,  
$56.0$ MeV fm$^{-3}$, and 
$94.92$ MeV fm$^{-3}$, respectively. 
In each set, the upper, middle and 
lower marks are for the cases 
of $R_{\infty}=8.2$ km,  $6.0$ km, and 
$3.8$ km, respectively. 
The labels in this figure correspond 
to the stellar model in Table \ref{tab_property_QS}.
For (b), the plot for $B=471.3$ MeV fm$^{-3}$ is also included
as the diamond.}
\label{fig_fmode}
\end{figure}
\end{center}

We plot the complex frequencies of the $f$-mode for $l=2$ in Fig. \ref{fig_fmode}. 
In this figure, the squares, triangles and circles denote 
the cases $B=28.9$ MeV fm$^{-3}$, 
$B=56.0$ MeV fm$^{-3}$ and 
$B=94.92$ MeV fm$^{-3}$, respectively. 
In each set, the upper, middle and lower marks correspond to 
the stellar models of $R_{\infty}=8.2$ km, 
$R_{\infty}=6.0$ km and $R_{\infty}=3.8$ km, respectively. 
The labels in this figure correspond to those of the 
stellar models in Table \ref{tab_property_QS}.

As can be seen in Fig. \ref{fig_fmode}(a), 
the $f$-mode frequency 
depends strongly on the bag constant but 
very weakly on the stellar radiation radius. This result agrees
with that of Kojima and Sakata \cite{Kojima2002}. Therefore, 
if the radiation radius of a quark star is determined by observation, 
even though the value of the radiation 
radius is somewhat uncertain, we can directly obtain the value of the bag 
constant by detecting $f$-mode QNMs (see Sec. \ref{sec:empiricalrelation}). 
Furthermore, we can restrict the 
mass of the source star by employing the observational value of the 
radiation radius.

The reason why the $f$-mode frequency depends strongly on the bag 
constant is understood as follows. The $f$-mode frequency 
depends strongly on average density 
but very weakly on the EOS \cite{Andersson1998}. 
In the case of quark stars with small radii, 
the average density is almost determined by the bag constant alone, 
as can be seen in Fig. \ref{fig_property}(c). Thus, 
the $f$-mode frequency depends strongly on the bag 
constant.

Moreover, we can see that the damping rate of the $f$-mode depends on 
the stellar radiation radius for each bag constant, which also agrees with
the result by Kojima and Sakata \cite{Kojima2002}. 
So, if we get 
data on the damping rate of the $f$-mode by observation, 
we can determine the stellar 
radiation radius independently of X-ray observations. 
However, one must keep in mind that the measurement of frequency is 
far more accurate than that of damping rate, 
because the estimated relative error in frequency is 
about three orders of magnitude smaller than that in damping 
rate \cite{Kokkotas2001}.

The $f$-mode frequency of a quark star with $M=0.7 M_{\odot}$ 
and $B=471.3$ MeV fm$^{-3}$ ($N1$) is plotted in
Fig. \ref{fig_fmode}(b). 
In this figure, results for the case $B \le 
B_{\text{max}}$ are also plotted for comparison. Compared with the 
$f$-mode frequencies for $B \le B_{\text{max}}$, 
the plot for $B=471.3$ 
MeV fm$^{-3}$ is relatively far away in phase space. 
This is mainly due to the large difference in the 
average density of quark stars.

\subsubsection{$w$- and $w_{\rm II}$-modes}

\begin{center}
\begin{figure}[htbp]
\includegraphics{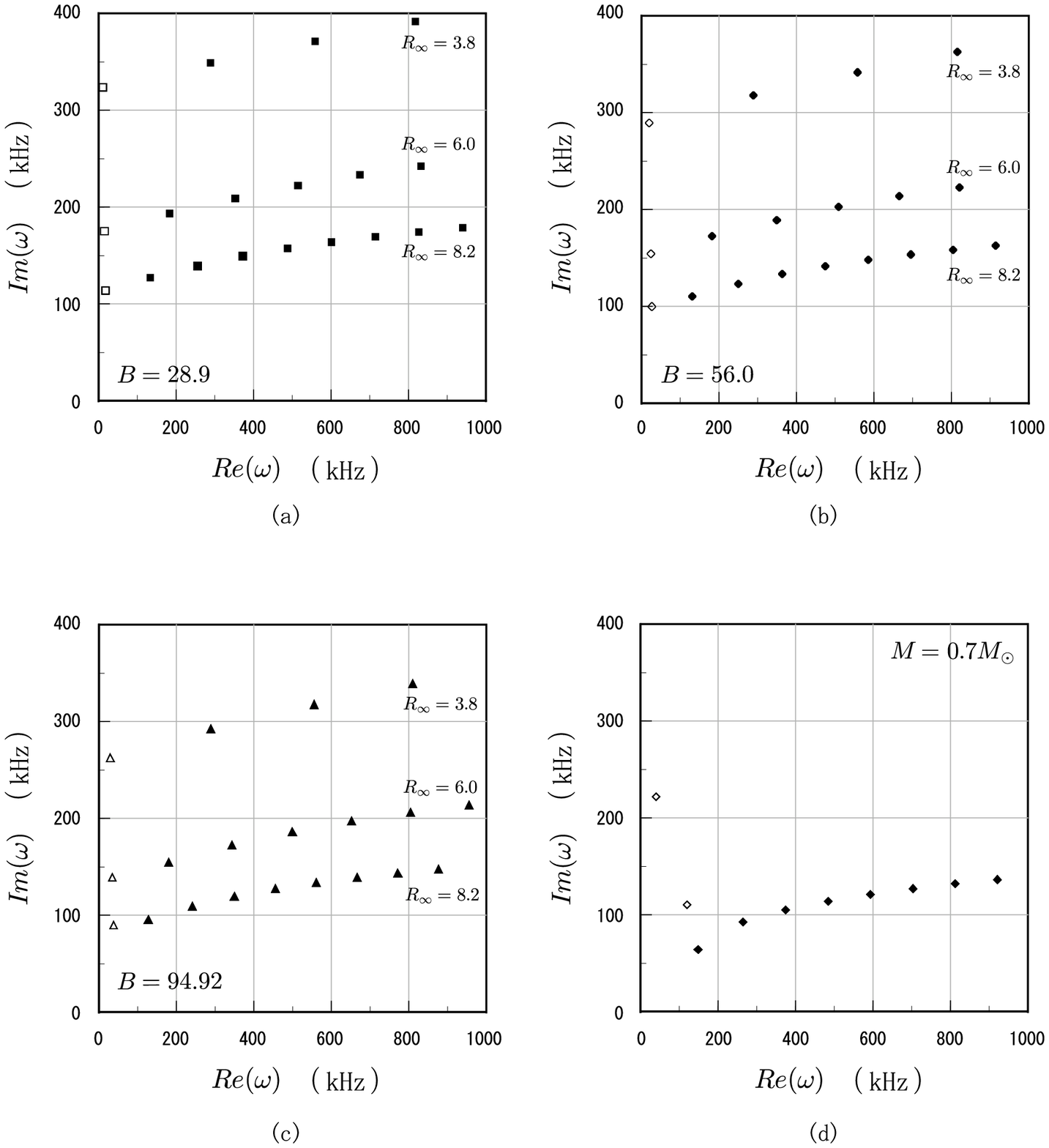}
\caption{Complex frequencies of both $w$- and 
$w_{\rm II}$-modes (a) for $B=28.9$ MeV fm$^{-3}$,
(b) for $B=56.0$ MeV fm$^{-3}$, 
(c) for $B=94.92$ MeV fm$^{-3}$, and 
(d) for $B=471.3$ MeV fm$^{-3}$.
For (a)-(c), the upper, middle and lower 
sequences correspond to the quark stars of $R_{\infty}=3.8$ km, 
$6.0$ km, and $8.2$ km, respectively. 
For (d), the mass of the stellar model is $0.7M_{\odot}$.
Filled and non-filled marks denote $w$- and $w_{\rm II}$-modes,
respectively.
These stellar models are listed in Table \ref{tab_property_QS}.}
\label{fig_wmode}
\end{figure}
\end{center}

The calculated complex frequencies of $w$- and 
$w_{\rm II}$-modes for $l=2$ 
are plotted in Fig. \ref{fig_wmode}. In 
these figures, the filled and non-filled marks correspond 
to $w$- and $w_{\rm II}$-modes, respectively. 
Figures \ref{fig_wmode}(a), (b) and (c) correspond to 
$B=28.9$, $56.0$ and $94.92$ MeV fm$^{-3}$, respectively. In each 
figure, the upper, middle and lower sequences correspond 
to the stellar models of 
$R_{\infty}=3.8$ km ($A1$, $B1$, $C1$), $R_{\infty}=6.0$ 
km ($A2$, $B2$, $C2$) and $R_{\infty}=8.2$ km ($A3$, $B3$, $C3$),
respectively.
In Fig. \ref{fig_wmode}(d), we plot the complex frequencies of $w$- and 
$w_{\rm II}$-modes of the star of $0.7 M_{\odot}$ 
with $B=471.3$ MeV fm$^{-3}$ ($N1$).

In order to obtain the tendency of $w$- and 
$w_{\rm II}$-modes with changing compactness, we 
calculate the $w$- and $w_{\rm II}$-modes of the stellar 
models whose compactness is much larger than that of the stellar 
models depicted in Figs. \ref{fig_wmode}(a), (b) and (c).
From this calculation, we find 
two $w_{\rm II}$-modes for the stellar model of higher 
compactness, for example, for $\rho_c = 1.4 \times 
10^{15}$ g/cm$^{-3}$ with $B = 56.0$ MeV fm$^{-3}$. In this case, the 
complex frequencies of $w$- and $w_{\rm II}$-modes 
appear similarly as in
Fig. \ref{fig_wmode}(d). 
Furthermore, we see that as the compactness of 
quark star gets smaller, the complex frequencies of the 
$w_{\rm II}$-modes shift toward the imaginary axis, 
while both the frequency and damping rate of $w$-modes 
become slightly larger. 
If the compactness is very small, 
the frequencies of $w_{\rm II}$-modes 
can be too small to find
except for the lowest $w_{\rm II}$-mode which has the largest frequency
among all $w_{\rm II}$-modes. 
That is why
Figs. \ref{fig_wmode}(a), (b) and (c) are quite different 
from the results obtained by Yip, Chu and Leung \cite{Yip1999} 
in the location of $w_{\rm II}$-modes.

\begin{center}
\begin{figure}[htbp]
\includegraphics{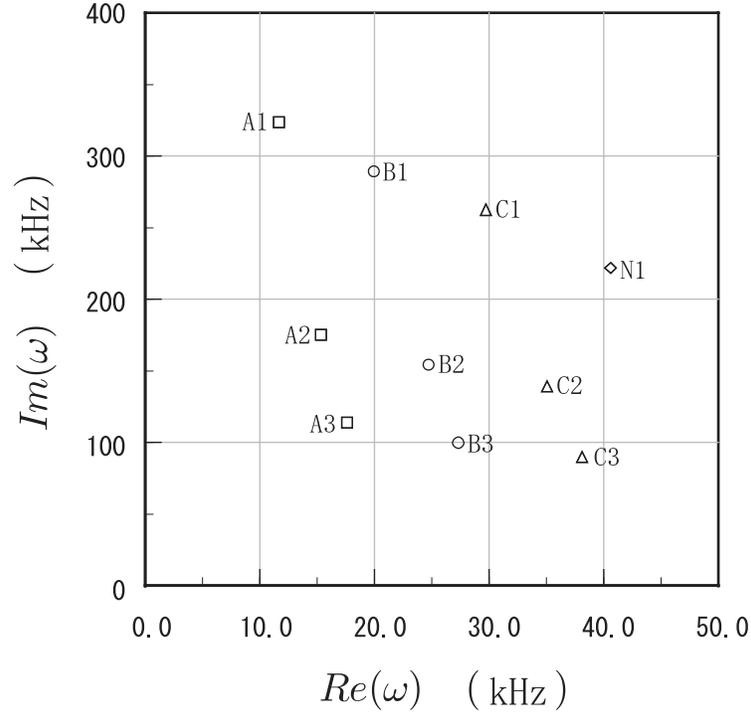}
\caption{Complex frequencies of the lowest $w_{\rm II}$-mode 
for each stellar model except for the plot for $B = 471.3 \mbox{MeV fm}^{-3}$, 
for which the second $w_{\rm II}$-mode is plotted.
The labels in this figure correspond to those 
of the stellar models in Table \ref{tab_property_QS}.}
\label{fig_wmode1}
\end{figure}
\end{center}

We pick up the lowest $w_{\rm II}$-mode and plot in 
Fig. \ref{fig_wmode1} for each stellar model. 
The frequency of the lowest $w_{\rm II}$-mode 
depends strongly on the bag constant. 
In this figure, the labels correspond to those of the 
stellar model in Table \ref{tab_property_QS}. Thus, we can
get information about the bag constant or the stellar 
radiation radius by observing the lowest $w_{\rm II}$-mode
frequency. 
Moreover, we could obtain a more stringent restriction
by using the damping rate,
although it seems more difficult to determine the 
damping rate by observation as Andersson and Kokkotas 
mentioned \cite{Andersson1996}.

On the other hand, in Fig. \ref{fig_wmode}, we see 
that both the frequency and damping rate of $w$-modes 
have little dependence on the bag constant. Therefore it is
difficult to determine the value of the bag constant by employing the 
observation of $w$-modes. In fact, the $w$-mode frequencies are 
all far beyond the frequency band for which 
gravitational wave 
interferometers such as LIGO, VIRGO, GEO600 and 
TAMA300 have good sensitivity \cite{Thorne1998}.

As seen in Fig. \ref{fig_wmode}(d), because the compactness of 
model $N1$ is much larger than that of the stellar models constructed 
with $B \le B_{\text{max}}$, the appearance of the complex frequencies
of the $w$- and $w_{\rm II}$-modes is quite different from that
for the case of $B \le B_{\text{max}}$. 
In Fig.\ref{fig_wmode1}, we also plot the $w_{\rm II}$-mode complex frequency
closest to the imaginary axis among the $w_{\rm II}$-modes previously found 
for model $N1$. 
From this result, we see that 
if the bag constant is too large, or is as large as the one adopted by Nakamura, 
it may be difficult to distinguish between 
the lowest $w_{\rm II}$-mode for the case of $B \le B_{\text{max}}$ 
and one which is closest to the imaginary axis 
for the case of $B=471.3$ MeV fm$^{-3}$.
In respect of this problem, even if we cannot distinguish between 
the lowest $w_{\rm II}$-mode for the case of $B \le B_{\text{max}}$ 
and one which is closest to the imaginary axis 
for the case of $B=471.3$ MeV fm$^{-3}$,  
it may be possible to discriminate between these two kinds of $w_{\rm II}$-modes 
by employing the observed $f$-mode frequency, 
because the plot of $f$-mode for $B=471.3$ 
MeV fm$^{-3}$ is relatively far away in phase space, compared with the 
$f$-mode frequencies for $B \le B_{\text{max}}$.

\section{Empirical formula}
\label{sec:empiricalrelation}

Kokkotas, Apostolatos and Andersson constructed the empirical formulas
for $f$-mode of  neutron stars \cite{Kokkotas2001};
\begin{eqnarray}
  \frac{Re(\omega)}{1\text{kHz}} 
  &&\cong  2\pi\,\left[(0.78\pm 0.01)+(1.63\pm 0.01)
  \left(\frac{M}{1.4M_{\odot}}\right)^{1/2}
  \left(\frac{10\text{km}}{R}\right)^{3/2}\right],  \label{empirical-real} \\
  \frac{Im(\omega)}{1\text{Hz}} 
  &&\cong  2\pi\,\left(\frac{M}{1.4M_{\odot}}\right)^3
  \left(\frac{10\text{km}}{R}\right)^4
  \left[(22.85\pm 1.51)-(14.65\pm 1.32)
  \left(\frac{M}{1.4M_{\odot}}\right)
  \left(\frac{10\text{km}}{R}\right)\right].  \label{empirical-imag}
\end{eqnarray}
We apply these empirical formulas to quark star models, and 
show the results in Table \ref{tab_empirical1}. 
As seen these results, 
though these formulas are deduced by employing various EOS including
the realistic one for neutron stars,  
it is found that a simple extrapolation to quark stars is not very successful. 
We try to construct the alternative formula for quark star models by using 
our numerical results for $f$-mode QNMs, but we can not do this using the same 
method as Kokkotas et al., because some special relation between $f$-mode 
QNMs and the stellar properties can not be found.

\begin{center}
\begin{figure}[htbp]
\includegraphics{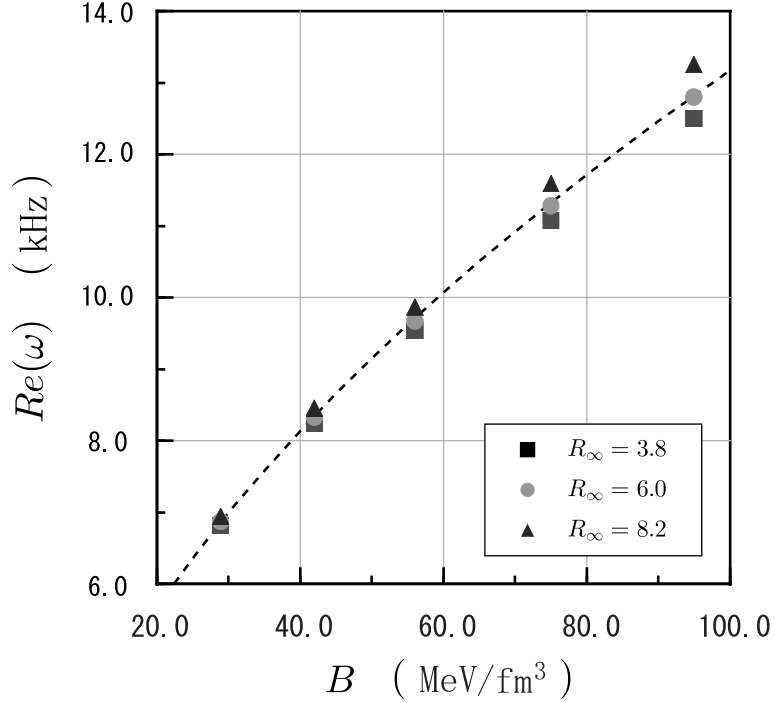}
\caption{The horizontal axis is the bag constant $B$ (MeV/fm$^{-3}$) 
and the vertical axis is the $f$-mode frequency $Re(\omega)$ .
The squares, circles and triangles correspond to the stellar models 
whose radiation radii are fixed $3.8$, $6.0$ and $8.2$ km, respectively. 
The dotted line denotes the new 
empirical formula (\ref{empirical2})
obtained in Sec. \ref{sec:empiricalrelation}
between $Re(\omega)$ of $f$-mode and $B$.}
\label{fig_empirical2}
\end{figure}
\end{center}

In the previous section, we see that the $f$-mode frequency depends strongly on 
not the stellar radiation radius but the bag constant. 
So, in order to get the relationship between $f$-mode frequency $Re(\omega)$ 
and the bag constant $B$, 
we calculate $f$-mode QNMs emitted by six more stellar models in addition to
the above nine stellar models (A1$\sim$A3, B1$\sim$B3, C1$\sim$C3). 
The radiation radii of these six added models are fixed at $R_{\infty}=3.8$, 
$6.0$ and $8.2$ km, and the adopted bag constants are $B=42.0$ and 
$75.0$ MeV fm$^{-3}$. The results of calculation are plotted in 
Fig. \ref{fig_empirical2}, and listed in table \ref{tab_empirical2}. 
In Fig. \ref{fig_empirical2}, squares, triangles and circles 
correspond to the stellar models 
whose radiation radii are fixed at $3.8$, $6.0$ and $8.2$ km, respectively. 
From our numerical results, we find the following new 
empirical formula between 
the $f$-mode frequency $Re(\omega)$ and the bag constant $B$: 
\begin{equation}
\frac{Re(\omega)}{1\text{kHz}}\,
 \approx 1.173\times \left(\frac{B}{1\text{MeV fm}^{-3}}\right)^{0.5252}.
 \label{empirical2}
\end{equation}
This empirical formula is also shown 
with a dotted line in Fig. \ref{fig_empirical2}.
In Table \ref{tab_empirical2}, we listed the bag constant 
which is obtained by substituting the given $f$-mode frequencies 
into the empirical formula (\ref{empirical2}), 
and its deviation from the true value. 
Here we calculate the deviation as
\begin{equation}
\frac{B-B_{emp}}{B},
\end{equation}
where $B$ and $B_{emp}$ are the bag constant we adopted for calculation of QNMs 
and the one given by substituting the $f$-mode frequency into the above 
empirical formula, 
respectively.
We see that this empirical formula is very useful, 
because it is possible to determine the bag constant precisely
by employing the observation of $f$-mode frequency,
even if the star has some range of radiation radii.

\section{Conclusion}
\label{sec:conclusion}

We calculate the nonradial oscillations of quark stars whose radiation 
radii are in the range of $3.8 \alt R_{\infty} \alt 8.2$ km, 
paying particular attention to the $f$-, $w$- and $w_{\rm II}$-modes of
quark stars constructed with the  bag model EOS.

We adopt four values for the bag constant and construct ten stellar models. 
We find that because the frequency of the $f$-mode strongly depends 
on the bag constant, we can restrict the bag constant by 
observations of the $f$-mode frequency. 
Moreover, since the damping rate of the 
$f$-mode depends on the stellar radius for each bag constant, 
we can restrict the stellar radius by detailed 
observations of the $f$-mode damping rate.

On the other hand, 
both the  frequency and damping rate of 
the lowest $w_{\rm II}$-mode also 
strongly depend on the bag constant and the stellar radius.
Therefore we may be able to 
get information about the properties of 
quark stars or the EOS governing quark 
matter from observations of the lowest $w_{\rm II}$-mode frequency.
If the damping rate of this mode were obtained,
we could derive more stringent constraints on the properties and 
EOS parameter.

Furthermore, we deduce a useful empirical formula between 
the frequency of $f$-mode and the bag constant. 
By using this relation, if the $f$-mode frequency is detected, 
we can determine the bag constant 
even if the radiation radius of the source star is not 
determined precisely.
Then we can determine the stellar radiation radius by employing 
the observation of the $f$-mode damping rate.

The dependence of the $f$-mode frequency
on the bag constant and radiation radius 
is different from that of 
the lowest $w_{\rm II}$-mode frequency.
Therefore, the lowest $w_{\rm II}$-mode QNM 
can help us to decide the bag constant and/or 
the stellar radiation radius.

We also calculate the $f$-, $w$- and $w_{\rm II}$-mode 
complex frequencies for a quark 
star of mass $M=0.7 M_{\odot}$ with $B=471.3$ MeV fm$^{-3}$. 
Because of the large
difference in compactness (or average density) of the 
quark star, the set of frequencies and damping rates of these 
QNMs differ considerably 
from that for quark stars with 
conventional values for $B$.
This implies that gravitational wave observations have strong
potential to test Nakamura's formation scenario for quark stars.
In the near future, we can get 
information about the EOS of quark matter by detecting gravitational 
waves from objects composed of quark matter.
Based on the information obtained, we can test the current 
understanding of quark matter and may obtain a 
more realistic picture of 
cold quark matter and nonperturbative QCD.

\acknowledgments

We would like to thank K. Maeda for useful discussion. 
We are also grateful to Y. Kojima, K. Kohri and 
O. James for valuable comments. 
This work was partly supported by the Grant-in-Aid
for Scientific Research (No. 05540) from the 
Japanese Ministry of Education, Culture, Sports, 
Science and Technology.



\begin{table}[htbp]
\caption{Properties of quark stars  constructed by the  bag model EOS.}
\label{tab_property_QS}
   \begin{center}
     \begin{tabular}{c|c|c|c|c|c|c}
        & $B\,\,(\text{MeV/fm}^3)$          & $R_{\infty}\,\,(\text{km})$ &
       $\rho_{c}\,\,(\text{g/cm}^3)$   & $M/M_{\odot}$ &
       $R\,\,(\text{km})$  & $M/R$\\  \hline

       $A1$ & $28.90$  &  $3.800$                         &  $2.090\times 10^{14}$
                     &  $2.329\times 10^{-2}$ & $3.765$ & 
$9.135\times 10^{-3}$   \\
       $A2$ & $28.90$  &  $6.000$                         &  $2.135\times 10^{14}$
                     &  $8.877\times 10^{-2}$  & $5.864$ & 
$2.235\times 10^{-2}$   \\
       $A3$ & $28.90$  &  $8.200$                         &  $2.205\times 10^{14}$
                     &  $2.165\times 10^{-1}$ & $7.859$   & 
$4.067\times 10^{-2}$   \\
       $B1$ & $56.00$  &  $3.800$                         &  $4.104\times 10^{14}$
                     &  $4.422\times 10^{-2}$ & $3.733$ & 
$1.749\times 10^{-2}$   \\
       $B2$ & $56.00$  &  $6.000$                         &  $4.285\times 10^{14}$
                     &  $1.637\times 10^{-1}$ & $5.742$ & 
$4.210\times 10^{-2}$   \\
       $B3$ & $56.00$  &  $8.200$                         &  $4.585\times 10^{14}$
                     &  $3.839\times 10^{-1}$ & $7.560$ & 
$7.498\times 10^{-2}$   \\
       $C1$ & $94.92$  &  $3.800$                         &  $7.095\times 10^{14}$
                     &  $7.283\times 10^{-2}$ & $3.688$ & 
$2.916\times 10^{-2}$   \\
       $C2$ & $94.92$  &  $6.000$                         &  $7.662\times 10^{14}$
                     &  $2.592\times 10^{-1}$ & $5.573$ & 
$6.867\times 10^{-2}$   \\
       $C3$ & $94.92$  &  $8.200$                         &  $8.730\times 10^{14}$
                     &  $5.776\times 10^{-1}$ & $7.156$ & 
$1.192\times 10^{-1}$  \\
       $N1$ & $471.3$  &  $5.661$ & $1.486\times 10^{16}$
                     &  $7.000\times 10^{-1}$ & $3.856$ & $2.681\times 10^{-1}$
     \end{tabular}
   \end{center}
\end{table}

\begin{table}[htbp]
\caption{Comparison between the $f$-mode QNMs of quark stars 
and value given by the empirical formulas
(\ref{empirical-real}) and (\ref{empirical-imag}). 
In this table, $Re(\omega)_{emp}$ and $Im(\omega)_{emp}$ 
express the range of frequency and damping rate respectively 
given by using the empirical formulas.}
\label{tab_empirical1}
   \begin{center}
     \begin{tabular}{c|c|c|c|c}
       & $Re(\omega)\,\,(\text{kHz})$ & $Im(\omega)\,\,(\text{Hz})$ & 
       $Re(\omega)_{emp}\,\,(\text{kHz})$  &  $Im(\omega)_{emp}\,\,(\text{Hz})$
       \\  \hline
       A1 & $6.822$  &  $2.98\times 10^{-3}$  & $10.52-10.72$ & 
                 $(2.97-3.42)\times 10^{-2}$ \\
       A2 & $6.872$  &  $2.67\times 10^{-2}$  & $10.55-10.74$ & 
                 $(2.66-3.10)\times 10^{-1}$ \\
       A3 & $6.945$  &  $1.12\times 10^{-1}$  & $10.58-10.78$ & 
                 $1.11-1.32$ \\
       B1 & $9.540$  &  $2.05\times 10^{-2}$  & $12.77-12.99$ & 
                 $(2.04-2.37)\times 10^{-1}$ \\
       B2 & $9.676$  &  $1.69\times 10^{-1}$  & $12.84-13.06$ & 
                 $1.67-2.00$ \\
       B3 & $9.873$  &  $6.40\times 10^{-1}$  & $12.95-13.17$ & 
                 $6.17-7.74$ \\
       C1 & $12.50$  &  $9.18\times 10^{-2}$  & $15.20-15.46$ & 
                 $0.913-1.07$ \\
       C2 & $12.80$  &  $6.83\times 10^{-1}$  & $15.36-15.62$ & 
                 $6.63-8.24$ \\
       C3 & $13.26$  &  $2.28$  & $15.64-15.90$ & 
                 $20.4-28.1$ \\
       N1 & $38.10$  &  $20.6$  & $34.90-35.39$ & $22.5-251.3$ 
     \end{tabular}
   \end{center}
\end{table}

\begin{table}[htbp]
\caption{Deviation from the new empirical formula (\ref{empirical2}) 
for each stellar model in Fig.\ref{fig_empirical2}. In this table, 
$B_{emp}$ is the value of the bag constant which is calculated by using 
the empirical formula (\ref{empirical2}) obtained 
in Sec. \ref{sec:empiricalrelation}.}
\label{tab_empirical2}
   \begin{center}
     \begin{tabular}{c|c|c|c|c}
       $R_{\infty}\,\,(\text{km})$ & $B\,\,(\text{MeV/fm}^3)$ & 
       $Re(\omega)\,\,(\text{kHz})$  &  $B_{emp}\,\,(\text{MeV/fm}^3)$
       & $deviation$\,\,(\%) \\  \hline
       $3.8$ & $28.90$  &  $6.822$  & $28.67$ & $-0.813$ \\
       $3.8$ & $42.00$  &  $8.242$  & $41.02$ & $-2.344$ \\
       $3.8$ & $56.00$  &  $9.540$  & $54.10$ & $-3.385$ \\
       $3.8$ & $75.00$  &  $11.076$  & $71.79$ & $-4.285$ \\
       $3.8$ & $94.92$  &  $12.503$  & $90.30$ & $-4.868$ \\
       $6.0$ & $28.90$  &  $6.872$  & $29.07$ & $0.581$ \\
       $6.0$ & $42.00$  &  $8.331$  & $41.85$ & $-0.355$ \\
       $6.0$ & $56.00$  &  $9.676$  & $55.57$ & $-0.768$ \\
       $6.0$ & $75.00$  &  $11.286$  & $74.39$ & $-0.819$ \\
       $6.0$ & $94.92$  &  $12.803$  & $94.44$ & $-0.503$ \\
       $8.2$ & $28.90$  &  $6.945$  & $29.65$ & $2.608$ \\
       $8.2$ & $42.00$  &  $8.458$  & $43.07$ & $2.554$ \\
       $8.2$ & $56.00$  &  $9.873$  & $57.74$ & $3.101$ \\
       $8.2$ & $75.00$  &  $11.597$  & $78.31$ & $4.415$ \\
       $8.2$ & $94.92$  &  $13.257$  & $100.89$ & $6.294$
     \end{tabular}
   \end{center}
\end{table}

\newpage

\end{document}